\documentclass[%
pof,%
 amsmath,amssymb,
 reprint,%
onecolumn,
]{revtex4-1}
\usepackage{CJKutf8}
\usepackage{graphicx}
\usepackage{dcolumn}
\usepackage{bm}

\begin{document}

\preprint{AIP/123-QED}

\begin{CJK}{UTF8}{}
\CJKfamily{mj}
\title{Impact of charge distribution of soft layers on transient electroosmotic flow of Maxwell fluids in soft nanochannels}

\author{Jun-Sik Sin(신준식)}
\email{js.sin@ryongnamsan.edu.kp}
\affiliation{Natural Science Center, \textbf{Kim Il Sung} University, Taesong District, Pyongyang, Democratic People's Republic of Korea}
\author{Nam-Il Ri(리남일)}
\affiliation{Natural Science Center, \textbf{Kim Il Sung} University, Taesong District, Pyongyang, Democratic People's Republic of Korea}
\author{Hyon-Chol Kim(김현철)}
\affiliation{Natural Science Center, \textbf{Kim Il Sung} University, Taesong District, Pyongyang, Democratic People's Republic of Korea}
\author{Sin-Hyok Hyon(현신혁)}
\affiliation{Natural Science Center, \textbf{Kim Il Sung} University, Taesong District, Pyongyang, Democratic People's Republic of Korea}


\begin{abstract}
\large
 
We theoretically study transient electroosmotic flow of general Maxwell fluids through polyelectrolyte grafted nanochannel with a layered distribution of charges.

By applying the method of Laplace transform, we semi-analytically obtain transient electroosmotic flow from Cauchy momentum equation and Maxwell constitutive equation.

 For nanochannels grafted with polyelectrolyte layers having different layered distribution of charges, we study the influence of dimensionless relaxation time, dimensionless polyelectrolyte layer thickness and dimensionless drag coefficient on transient electroosmotic flow. 

We present the results for some particular cases. 
Firstly, we unravel that for the case of polyzwitterionic brush that the sum of positive and negative structural charges is zero, total electroosmotic flow is non-zero. In particular, depending on charge distribution within end part of polyelectrolyte layers, the direction of electroosmotic flow can be reversed critically.
Secondly, in order to quantitatively evaluate a reversal of electroosmotic flow for two polyelectrolyte layers of opposite signs, we introduce a critical number $k_s$ as the ratio between layered charge densities of two polyelectrolye layers. Increasing $k_s$ allows electroosmotic flow to be reversed easily.

We verify that adjusting charge distributions of the layer can control intentionally the direction of the flows as well as strength of electroosmotic flow.

\end{abstract}
\maketitle
\end{CJK}

\large

\section{Introduction}

Microfluidics technology opens a widespread avenue for developing microchips which can support analysis of medical and biochemical samples. \cite%
{eijkel_2009,takehara_2002, khatibi_jml_2022, khatibi_lang_2022, khatibi_an_2023, khatibi_pf_2022, zhu_2018,jiang_2018,qian_2016,jiang_2019,davis_2017,karimi_2017, ghosh_2020}

A variety of routes to microchip design are currently being exploited to control the shape and direction of liquid flow. An important branch among them is based on the paradigm of electroosmosis that is triggered due to interaction of the electric double layer near a charged channel wall in an electrolyte solution with an external electric field parallel to the channel axis.

In particular, it was reported that the direction of electroosmosis flow can be sensitively altered by controlling electric field strength 
\cite%
{ewing_1993,li_2015}, surface patterning \cite%
{whitesides_2000}, polyelectrolyte multilayers \cite%
{locascio_2000}. 

Many researchers \cite%
{beskok_2019,kundu_2016,chakraborty_2015,pinho_2000,pinho_2016,ganguly_2017,alves_2013,sin_2018,yang_2009,liu_2012, liu_2013, wei_2013, chakraborty_2016, mendez_2016,chakraborty_2016_2,liu_2010,jian_2011}  focused their attention on favourable properties of direct current electroosmotic flow for numerous types of geometries, boundary conditions and types of fluids in microchannel.

Most of reported studies revealed that start-up electroosmotic flow is crucial for not only mixing of various species of liquids but also stability of microchips. Stratup flows can quite increase heat flow rate and fluid flow velocity and destroy microchip unit of the devices. Zhao and Yang \cite%
{yang_2009} determined analytical solutions for generalized Oldroyd-B flow under small symmetric zeta potential at the wall.
Many of studies using Newtonian, Phan-Thien-Tanner and Maxwell fluids confirmed that asymmetry of zeta potential can facilitate control of velocity distribution of electroosmotic flow. Jimenez et al \cite%
{mendez_2016} investigated transient electroosmotic flow of Maxwell liquid in rectangular microchannel under high asymmetric zeta potential by using semianalytical methods. Suman et al \cite%
{chakraborty_2016_2} probed startup flow of Oldroyd-B fluids in a microchannel with a consideration of interfacial charge and viscoelasticity of liquids.  Jian et al \cite%
{ liu_2010, jian_2011} determined semi-analytical solution of time periodic electro-osmotic flow through a microannulus and a rectangular microchannel.

On the other hand, it is well known that polyelectrolyte grafted nanochannels (so-called soft nanochannels) possess superior properties compared to rigid nanochannel. Ohshima et al \cite%
{ohshima_2015} studied a steady flow combined electroosmotic and pressure driven flow in soft nanochannel with polyelectrolyte brush charged with a uniform charge density under slip condition.

The authors of \cite%
{das_2017} studied a steady electrokinetic flow in end-charged polyelectrolyte grafted nanochannel and suggested that at medium and high salt concentrations, streaming potential and electrochemomechanical energy conversion efficiency can be increased. They \cite%
{das_2019,das_2020,das_2020_2} also probed electrokinetic phenomena in nanochannels grafted with pH-responsive polyelectrolyte brushes with the help of augmented strong stretching theory. The authors of \cite%
{sadeghi_2018,sadeghi_2019} investigated electroosmotic flow in soft nanochannel with variational principle.  In addition, researchers investigated the influences of various geometries \cite%
{khatibi_2020, khatibi_2021, hsu_2021}, ionic size \cite%
{pccp_2018} and porous permeable film \cite%
{vinogradova_2020} on elelctroosmotic flow through polyelectrolyte grafted nanochannel.

  The authors of \cite%
{liu_2017, yang_2016, yang_2017}  studied alternative current electrokinetics in uniform charged polyelectrolyte grafted nanochannel and provided analytical solutions for streaming potential and electrokinetic energy conversion efficiency through the nanochannel. 

The author of \cite%
{ohshima_2022} provided ion partitioning effect on solute transport through polyelectrolyte grafted nanochannel of fractional Jeffrey fluid through polyelectrolyte-coated nanopore with reversible wall reaction.

On the other hand, the authors of \cite%
{duval_2009, duval_2010, duval_2014, duval_2014_2,duval_2016} developed electrostatics and electrokinetics of polyelectrolyte grafted interface with layered distribution of charges. It is shown that charge-stratified systems can reverse their electrophoretic mobility and streaming current in monovalent electrolyte solution.

Recently, the authors of \cite{khatibi_pf_2023, ashrafi_ces_2022} investigated a hybrid nanochannel with a soft layer with different polyelectrolyte charge distirbutions like constant, exponential, sigmoidal and soft-step like functions. They demonstrated that manipulating the charge distribution can control ionic selectivity and consequently yield functionalized nanochannels for applying in separation, diagnostics, and sensing. However, the approach didn't consider transient electroosmotic flow and any complex charge density distribution of soft layer.

 Unlike previous studies, we for the first time study the electroosmotic flow in nanochannel with different charge distribtuion of soft layer by using inverse Laplace method. Our approach is semi-analytical and can be applied for analyzing transient electroosmosis through soft nanochannels and predicting reversal flow of electroosmosis.

The paper is organized as follows.

In the paper, we numerically analyze the electroosmotically driven transport of general Maxwell fluids through a nanochannel grafted with polyelectrolyte layers with different charge distributions.
First, Debye-H\"uckel approximation is applied to determine analytical solution for electric potential in nanochannels grafted with polyelectrolyte layers with different layered distribution of charges. Here, we assume that any charge distribution can be approximated as a layered-charge distribution and the thickness of polyelectrolyte grafted layer is constant. Second, the method of Laplace transform is used for obtaining semi-analytically transient electroosmotic velocity. Then, in different cases, we make discussion of startup flow through polyelectrolyte grafted nanochannels. Last but not least, the controllability of the direction and velocity of electroosmotic flow is predicted and investigated in more detail.

\begin{figure}
\begin{center}
\includegraphics[width=0.9\textwidth]{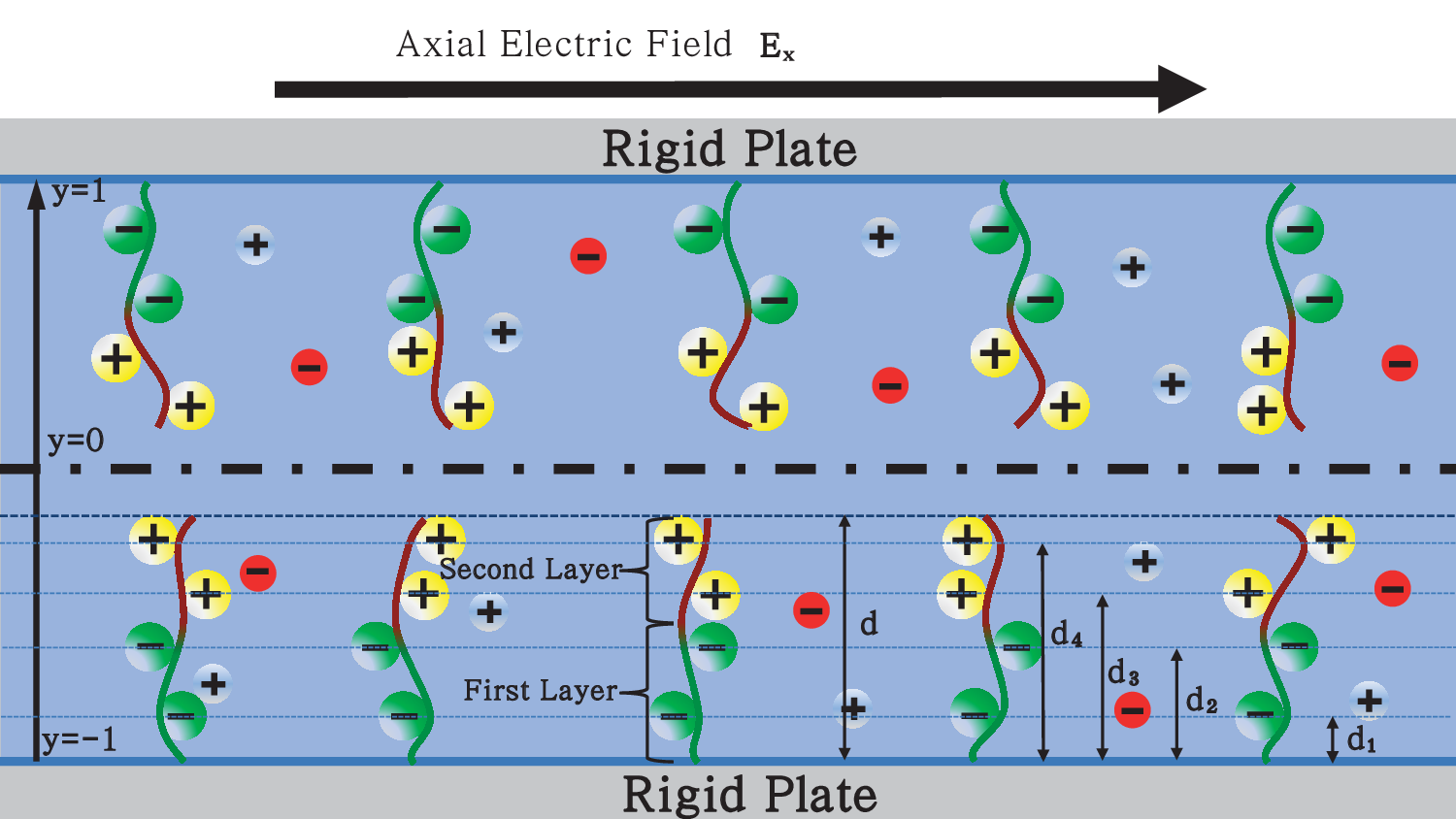}
\caption{(Color online) Schematic of transient electroosmotic flow of the generalized Maxwell fluids through a polyelectrolyte grafted nanochannel with a layered distribution of charges.  The electrolyte solution contains cations (cyan circles),  anions (red circles), polyelectrolyte anion (green circles) and polyelectrolyte cation (yellow circles).}
\label{fig:1}
\end{center}
\end{figure}

\section{Theory}

Fig. \ref{fig:1} shows schematics of transient electroosmotic flow of incompressible general Maxwell fluids through a nanochannel with the thickness of $2H$. We introduce 2-dimensional coordinate with Y-axis being perpendicular to a charged wall and X-axis being parallel to the wall. Top and bottom plates are located at $y^*=H$ and $y^*=-H$. $\varepsilon_0$ is the absolute permittivity of vaccum, $\varepsilon$ is the relative permittivity of the fluid, $k_B$ is the Boltzmann constant, $T$ is the temperature of the fluid, $z$ is the valence of the electrolyte, $e$ is the elementary charge of the electron, $n_0$ is the bulk ionic number density of electrolyte solution,  $E_x$ is the externally applied electric field parallel to the flow direction. 

We assume that the polyelectrolyte layer contains $n_p$ structural layered charges.
The surface charge density of layered charges is denoted as $\sigma_k  \left(k=1\sim n_p\right)$, where subscript $k$ means a number of the corresponding layer of charges.The distance from the nanochannel wall to $k$-th layer of charges is expressed as $d_k^*$.

\subsection{Electric potential distribution in nanochannel.}

The paper is based on the assumption that the electric potential of the non-uniform charge distribution of the polyelectrolyte layer can be expressed with respect to that of layered distribution of charges.
The principle of superposition of electric fields says that the total electric field at any position P equals to the vector sum of the fields at P due to each layered charge in the charge distribution.
Here, we assume that the electric field associated with each layered charge is sufficiently low.

 Therefore, Debye-H\"uckel approximation can be applied to Poisson-Boltzmann equation, Debye-H\"uckel equation is linear differential equation
 \begin{equation}
\frac{{d^2 \phi }}{{dy^2 }} = \frac{\phi }{{\lambda ^2 }},
\label{eq:1}
\end{equation}
where we use  $y = \frac{{y^* }}{H}$, $\lambda {\rm{ = }}\sqrt {\frac{{\varepsilon \varepsilon _0 k_B T}}{{2z^2 e^2 n_0}}}$ and  $\phi  = \frac{{ez\psi }}{{k_B T}}$.

Boundary conditions are as follows
 \begin{equation}
\frac{{d\phi }}{{dy}}\left| {_{y = 0} } \right. = 0,\,\frac{{d\phi }}{{dy}}\left| {_{y =  - 1} } \right. = 0,
\label{eq:201}
\end{equation}
\begin{equation}
 \frac{{d\phi }}{{dy}}\left| {_{y = \left( { - 1 + d_k } \right)_+  } } \right. - \frac{{d\phi }}{{dy}}\left| {_{y = \left( { - 1 + d_k } \right)_-  } } \right. = \frac{{ez\sigma _k H}}{{k_B T\varepsilon _0 \varepsilon }},\\
\phi \left| {_{y = \left( { - 1 + d_k } \right)_+  } } \right. = \phi \left| {_{y = \left( { - 1 + d_k } \right)_ -  } } \right. \quad \left(k=1
 \sim n_p\right),
\label{eq:202}
\end{equation}
where $\sigma_k$ is the surface charge density of k-th layer of charges, $d = \frac{{d^* }}{H}$ and $d_k  = \frac{{d_k^* }}{H}$.
Subscript - refers to the region $-1 \le y \le -1+d_k$, and subscript + refers to the region $-1+d_k \le y \le 0 $.

The first expression of Eq.(\ref{eq:201}) means that the hard part of nanochannel wall is uncharged.  The second one indicates that at the middle point of nanochannel, the electric field strength is zero, attributed to geometrical symmetry of nanochannel. Eq. (\ref{eq:202}) represents continuous condition of electric field strength and electrostatic potential at the postion of k-th layered charge.
Since the Debye-H\"uckel equation is a linear differential equation, we anticipate that the self-consistent electric fields due to electrolyte ions and the total layered charge system on polyelectrolyte chains in an electrolyte solution equals to the vector sum of the self-consistent electric fields due to electrolyte ions and the individual layered charges on polyelectrolyte chains.

We denote the electrostatic potential induced by $k$-th layered charge as $\phi_k$.

In the region of ($ - 1 + d_k  \le y \le 0$)
\begin{equation}
\phi _k  = \frac{{ez\sigma _k \lambda H}}{{k_B T\varepsilon _0 \varepsilon }}\frac{{\cosh \left( {\frac{{d_k }}{\lambda }} \right)}}{{\sinh \left( {\frac{1}{\lambda }} \right)}}\cosh \left( {\frac{y}{\lambda }} \right),
\label{eq:3}
\end{equation}

In the region of ($- 1 \le y \le  - 1 + d_k$)
 \begin{equation}
\phi _k  = \frac{{ez\sigma _k \lambda H}}{{k_B T\varepsilon _0 \varepsilon }}\frac{{\cosh \left( {\frac{{1 - d_k }}{\lambda }} \right)}}{{\sinh \left( {\frac{1}{\lambda }} \right)}}\cosh \left( {\frac{{1 + y}}{\lambda }} \right).
\label{eq:4}
\end{equation}

The total electric potential can be expressed as
\begin{eqnarray}
 \phi  = \sum\limits_k {\phi _k \left( y \right)}  =\frac{{ez\lambda H}}{{k_B T\varepsilon _0 \varepsilon \sinh \left( {\frac{1}{\lambda }} \right)}}\sum\limits_k {\sigma _k \cosh \left( {\frac{{d_k }}{\lambda }} \right)\cosh \left( {\frac{y}{\lambda }} \right)\Theta \left( {y + 1 - d_k } \right)} + \\ \nonumber
+ \frac{{ez\lambda H}}{{k_B T\varepsilon _0 \varepsilon \sinh \left( {\frac{1}{\lambda }} \right)}}\sum\limits_k {\sigma _k \cosh \left( {\frac{{1 - d_k }}{\lambda }} \right)\cosh \left( {\frac{{1 + y}}{\lambda }} \right)\Theta \left( { - 1 + d_k  - y} \right)},
\label{eq:5}
 \end{eqnarray}
where $\Theta\left(y\right)$ is  Heaviside  step function.
We already know that if each of two functions satisfies a linear differential equation, a linear combination of the two functions also becomes as a solution of the differential equation. Furthermore, considering boundary conditions, the linear combination is limited to the case of sum of the two functions. We can easily verify that the electric potential satisfies Eq. (\ref{eq:1}) and boundary conditions Eqs. (\ref{eq:201}, \ref{eq:202}).

\subsection{Transient electroosmotic flow of general Maxwell fluids through nanochannel}
Considering pure electroosmotic flow without pressure gradient, Cauchy momentum can be written as follows
 \begin{equation}
\rho \frac{{\partial u^{\rm{*}} }}{{\partial t^{\rm{*}} }} = \frac{{\partial \tau _{yx} }}{{\partial y^{\rm{*}} }} + \rho _e E_x  - \mu _c u^{\rm{*}} ,\,\,\,\,\,\,\,\,\,\,\,\,\,\,\,\,\, - H \le y^{\rm{*}}  \le  - H + d^{\rm{*}} 
\label{eq:6}
\end{equation}
\begin{equation}
\rho \frac{{\partial u^* }}{{\partial t^* }} = \frac{{\partial \tau _{yx} }}{{\partial y^* }} + \rho _e E_x ,\,\,\,\,\,\,\,\,\,\,\,\,\, - H + d^*  \le y^*  \le 0
\label{eq:7}
\end{equation}
where $u^*\left(y^*, t^*\right)$ is the velocity along X-axis,  $\rho$ is the volume density of fluids,  $t^*$ is the time, $\tau_{yx}$ is stress tensor, $\rho_e \left(y\right)$ is the volume charge density and $\mu_c$ is the drag coefficient within polyelectrolyte layer. In fact, we assume that transient electroosmotic flow doesn't affect the charge distribution inside Debye layer. 
For a general Maxwell fluid, constitutive equation satisfies the following relation
 \begin{equation}
\left( {1 + \lambda _1 \frac{\partial }{{\partial t^* }}} \right)\tau _{yx}  = \mu \frac{{\partial u^*}}{{\partial y^* }},
\label{eq:8}
\end{equation}
where $\lambda_1$ is relaxation time and $\mu$ viscosity at zero shear rate.
Combining Eq. (\ref{eq:6}) and Eq. (\ref{eq:8}) provides the following equation 
 \begin{equation}
\rho \left( {1 + \lambda _1 \frac{\partial }{{\partial t^* }}} \right)\frac{{\partial u^* }}{{\partial t^* }} = \mu \frac{{\partial ^2 u^* }}{{\partial{ y^*} ^2 }} + \rho _e E_x  - \mu _c \left( {1 + \lambda _1 \frac{\partial }{{\partial t^* }}} \right)u^* .
\label{eq:9}
\end{equation}

Considering $\rho _e \left( y \right) =  - 2n_0 ze\sinh \left( \phi  \right) \approx  - 2n_0 ze\phi \left( y \right)$, Eq. (\ref{eq:9}) is reduced to 
\begin{equation}
\mu \frac{{{\partial ^2} u^* }}{{\partial {y^*} ^2 }} - \rho \lambda _1 \frac{{\partial ^2 u^* }}{{\partial{ t^*} ^2 }} - \left( {\rho  + \mu _c \lambda _1 } \right)\frac{{\partial u^* }}{{\partial t^* }} - \mu _c u^*  - 2n_0 ze\phi \left( y \right)E_x  = 0.
\label{eq:10}
\end{equation}
In the present study, the following boundary conditions are used for determining fluid flow.
 \begin{equation}
u\left| {_{y =  - 1} } \right. = 0,\,\,\,\,\frac{{du}}{{dy}}\left| {_{y = 0} } \right. = 0,\frac{{du}}{{dy}}\left| {_{y = \left( { - 1 + d} \right)_ +  } } \right. = \frac{{du}}{{dy}}\left| {_{y = \left( { - 1 + d} \right)_-  } } \right.,u\left| {_{y = \left( { - 1 + d} \right)_-  } } \right. = u\left| {_{y = \left( { - 1 + d} \right)_+  } } \right..
\label{eq:11}
\end{equation}
It should be emphasized that at all positions of layered charges, Cauchy   momentum equations are equal to each other and consequently boundary conditions at the positions are not necessary.
We introduce the following dimensionless quantities
\begin{equation}
u\left( {y,t} \right) = \frac{{u^* \left( {y,t} \right)}}{{U_0 }},{\kern 1pt} {\kern 1pt} \bar \tau _{yx}  = \frac{{\tau _{yx} }}{{\mu U_0 /H}},\bar \mu _c  = \frac{{\mu _c }}{{\mu /H^2 }}, t = \frac{{t^* }}{{\rho H^2 /\mu }},\bar \lambda _1  = \frac{{\lambda _1 }}{{\rho H^2 /\mu }},
\label{eq:13}
\end{equation}
from which Eq. (\ref{eq:10}) is rearranged as
 \begin{equation}
\frac{{\partial ^2 u}}{{\partial y^2 }} - \bar \lambda _1 \frac{{\partial ^2 u}}{{\partial t^2 }} - \left( {1 + \bar \mu _c \bar \lambda _1 } \right)\frac{{\partial u}}{{\partial t}} - \bar \mu _c u - 2n_0 ze\phi \left( y \right)\frac{{E_x H^2 }}{{\mu U_0 }} = 0.
\label{eq:14}
\end{equation}

 Let's make the method of Laplace transform as
\begin{equation}
U\left( {y,s} \right) = L\left[ {u\left( {y,t} \right)} \right] = \int_0^\infty  {u\left( {y,t} \right)e^{ - st} dt}. 
\label{eq:15}
\end{equation}

Considering the fact that at initial time flow velocity is zero at any position, $u\left( {y,0} \right) = 0$, Eq.(\ref{eq:14}) and Eq.(\ref{eq:7}) are transformed into
\begin{equation}
\frac{{d ^2 U\left( y \right)}}{{dy^2 }} - \left( {\bar \lambda _1 s + 1} \right)\left( {s + \bar \mu _c } \right)U\left( y \right) - \frac{{2n_0 ze\phi \left( y \right)E_x H^2 }}{{s\mu U_0 }} = 0
\label{eq:16}
\end{equation}
and 
\begin{equation}
\frac{{d ^2 U\left( y \right)}}{{dy^2 }} - \left( {\bar \lambda _1 s + 1} \right)s U\left( y \right) - \frac{{2n_0 ze\phi \left( y \right)E_x H^2 }}{{s\mu U_0 }} = 0.
\label{eq:161}
\end{equation}

Here, we again can apply the principle of superposition to obtain the whole solution. In other words, we first solve Eq. (\ref{eq:16} ) for  each item  $\phi _k \left( y \right)$ consisting of net electric potential distribution(see Eq.(\ref{eq:4})) and then summation of the individual solutions becomes as the solution of the original equation for the net electric potential distribution. 
The transient velocity solution for the $k$-th  term in Eq. (\ref{eq:16}) can be obtained by the following procedure.

$U_0$ and $\phi$ of Eq.(\ref{eq:16}) and Eq. (\ref{eq:161}) should be replaced as $U_{0k}$ and $\phi_k$, respectively.
Substituting $U_{0k}  = E_x \lambda H\sigma_k /\mu$ and Eqs. (\ref{eq:3}) and Eq.(\ref{eq:4}) into Eq.(\ref{eq:16}) and  Eq.(\ref{eq:161}), we obtain
\begin{equation}
 \frac{{d ^2 U_k\left( y \right)}}{{dy^2 }} - \beta _1 ^2 U_k\left( y \right) = \frac{{\frac{1}{{\lambda ^2 }}}}{s}\frac{{\cosh \left( {\frac{{d_k }}{\lambda }} \right)}}{{\sinh \left( {\frac{1}{\lambda }} \right)}}\cosh \left( {\frac{y}{\lambda }} \right),   \quad  \quad \left(-1+d\le y\le 0\right)
\label{eq:18}
\end{equation}
\begin{equation}
 \frac{{d ^2 U_k\left( y \right)}}{{dy^2 }} - \beta _2 ^2 U_k\left( y \right) = \frac{{\frac{1}{{\lambda ^2 }}}}{s}\frac{{\cosh \left( {\frac{{d_k }}{\lambda }} \right)}}{{\sinh \left( {\frac{1}{\lambda }} \right)}}\cosh \left( {\frac{y}{\lambda }} \right),   \quad  \quad \left(-1+d_k\le y\le -1+d\right)
\label{eq:181}
\end{equation}
\begin{equation}
 \frac{{d ^2 U_k\left( y \right)}}{{dy^2 }} - \beta _2 ^2 U_k\left( y \right) = \frac{{\frac{1}{{\lambda ^2 }}}}{s}\frac{{\cosh \left( {\frac{{1 - d_k }}{\lambda }} \right)}}{{\sinh \left( {\frac{1}{\lambda }} \right)}}\cosh \left( {\frac{{1 + y}}{\lambda }} \right),   \quad  \quad \left(-1\le y\le -1+d_k\right)
\label{eq:182}
\end{equation}
where $\beta _1 ^2  = \left( {\bar \lambda _1 s + 1} \right)s$ and $\beta _2 ^2  = \left( {\bar \lambda _1 s + 1} \right)\left( {s + \bar \mu _c } \right)$.

It is the principle of differential equation theory that the general solution of  an inhomogeneous differential equation should be obtained as sum of the general solution of the corresponding homogeneous differential equation and the particular solution of the inhomogeneous equation. 
\begin{equation}
U_k\left( {y,s} \right) = U_{kh} \left( {y,s} \right) + U_{ks} \left( {y,s} \right),
\label{eq:19}
\end{equation}
where $U_k\left( {y,s} \right)$, $U_{kh} \left( {y,s} \right)$ and $U_{ks} \left( {y,s} \right)$ denote the general  solution of the inhomogeneous differential equation, the  general solution of the corresponding homogeneous  equation and the particular solution of the inhomogeneous equation, respectively.
The general solution of the corresponding homogeneous equation can be written as 
\begin{equation}
U_{kh} \left( {y,s} \right) = C_1 \cosh \left( {\beta _1 y} \right) + C_2 \sinh \left( {\beta _1 y} \right).
\label{eq:20}
\end{equation}

The particular solution of the inhomogeneous equation is expressed as
\begin{equation}
U_{ks} \left( {y,s} \right) = C_3 \cosh \left( {\frac{y}{\lambda }} \right).
\label{eq:21}
\end{equation}
Substituting Eq.(\ref{eq:21}) into Eq.(\ref{eq:18}) gives
 \begin{equation}
C_3  = \frac{{\frac{1}{{\lambda ^2 }}}}{s}\frac{{\cosh \left( {\frac{{d_k }}{\lambda }} \right)}}{{\sinh \left( {\frac{1}{\lambda }} \right)}}/\left( {\frac{1}{{\lambda ^2 }} - \beta _1 ^2 } \right).
\label{eq:22}
\end{equation}

As in the region of ($-1+d \le y\le 0$), in the region of  ($ - 1+d_k \le y \le  - 1 + d $), the general solution of the corresponding homogeneous equation for Eq. (\ref{eq:181}) should be expressed as
   \begin{equation}
U_{kh} \left( {y,s} \right) = C_5 \cosh \left( {\beta _2 y} \right) + C_6 \sinh \left( {\beta _2 y} \right).
\label{eq:26}
\end{equation}
The particular solution of Eq. (\ref{eq:181}) should be determined by
\begin{equation}
U_{ks} \left( {y,s} \right) = C_7 \cosh \left( {\frac{y}{\lambda }} \right).
\label{eq:27}
\end{equation}

Then we can get the following expression for $C_7$
 \begin{equation}
C_7  = \frac{{\frac{1}{{\lambda ^2 }}}}{s}\frac{{\cosh \left( {\frac{{d_k }}{\lambda }} \right)}}{{\sinh \left( {\frac{1}{\lambda }} \right)}}/\left( {\frac{1}{{\lambda ^2 }} - \beta _2 ^2 } \right).
\label{eq:28}
\end{equation}
In the region of  ($ - 1 \le y \le  - 1 + d_k $), we get 
\begin{equation}
U_{kh} \left( {y,s} \right) = D_1 \cosh \left( {\beta _2 y} \right) + D_2 \sinh \left( {\beta _2 y} \right),
\label{eq:31}
\end{equation}
\begin{equation}
U_{ks} \left( {y,s} \right) = D_3 \cosh \left( {\frac{{1 + y}}{\lambda }} \right),
\label{eq:32}
\end{equation}
\begin{equation}
D_3  = \frac{{\frac{1}{{\lambda ^2 }}}}{s}\frac{{\cosh \left( {\frac{{1 - d_k }}{\lambda }} \right)}}{{\sinh \left( {\frac{1}{\lambda }} \right)}}/\left( {\frac{1}{{\lambda ^2 }} - \beta _2 ^2 } \right).
\label{eq:33}
\end{equation}
Considering boundary conditions of Eq. (\ref{eq:11}), we have the following expressions from which we can determine the coefficients $C_1, C_2, C_5, C_6, D_1$ and $D_2$  and consequently Laplace transform of velocity profile is found.
\begin{eqnarray}
\small
C_2  = 0,\\
\beta _1 C_1 \sinh \left( {\beta _1 \left( { - 1 + d} \right)} \right) + \frac{1}{\lambda }C_3 \sinh \left( {\frac{{ - 1 + d}}{\lambda }} \right) =\\\nonumber
= \beta _2 C_5 \sinh \left( {\beta _2 \left( { - 1 + d} \right)} \right) + \beta _2 C_6 \cosh \left( {\beta _2 \left( { - 1 + d} \right)} \right) + \frac{1}{\lambda }C_7 \sinh \left( {\frac{{ - 1 + d}}{\lambda }} \right),\\
C_1 \cosh \left( {\beta _1 \left( { - 1 + d} \right)} \right) + C_3 \cosh \left( {\frac{{ - 1 + d}}{\lambda }} \right) =\\\nonumber
= C_5 \cosh \left( {\beta _2 \left( { - 1 + d} \right)} \right) + C_6 \sinh \left( {\beta _2 \left( { - 1 + d} \right)} \right) + C_7 \cosh \left( {\frac{{ - 1 + d}}{\lambda }} \right),\\
\beta _2 C_5 \sinh \left( {\beta _2 \left( { - 1 + d_k } \right)} \right) + \beta _2 C_6 \cosh \left( {\beta _2 \left( { - 1 + d_k } \right)} \right) + \frac{1}{\lambda }C_7 \sinh \left( {\frac{{ - 1 + d_k }}{\lambda }} \right) =\\\nonumber
= \beta _2 D_1 \sinh \left( {\beta _2 \left( { - 1 + d_k } \right)} \right) + \beta _2 D_2 \cosh \left( {\beta _2 \left( { - 1 + d_k } \right)} \right) + \frac{1}{\lambda }D_3 \sinh \left( {\frac{{d_k }}{\lambda }} \right),\\
C_5 \cosh \left( {\beta _2 \left( { - 1 + d_k } \right)} \right) + C_6 \sinh \left( {\beta _2 \left( { - 1 + d_k } \right)} \right) + C_7 \cosh \left( {\frac{{ - 1 + d_k }}{\lambda }} \right) =\\\nonumber
= D_1 \cosh \left( {\beta _2 \left( { - 1 + d_k } \right)} \right) + D_2 \sinh \left( {\beta _2 \left( { - 1 + d_k } \right)} \right) + D_3 \cosh \left( {\frac{{d_k }}{\lambda }} \right),\\
D_1 \cosh \left( {\beta _2 } \right) - D_2 \sinh \left( {\beta _2 } \right) + D_3  = 0.
\end{eqnarray}

Although we can solve the equations, the solutions are too complex to show them explicitly. 

Once we get the Laplace transform of the transient velocity profile due to each layered-charge, considering the difference in normalization factor $U_{0k}$ between structural layered charges, all the solutions should be added  in order to get the Laplace transform of the overall transient electroosmotic velocity. Finally, we should make inverse Laplace transform to determine the transient velocity profile. To do this, we use the efficient approach for the numerical inversion of Laplace transform suggested by Zhao \cite%
{zhao_2004}. 

It is worthwhile that we consider only the regime of weak electrostatics because in the regime, the linearized Poisson-Boltzmann equation is valid and consequently semi-analytical formulae are obtained. Considering the above fact, our present theory can be applied to the case $exp\left(\phi\right)\approx \phi$, $\phi<<1$.

As demonstrated in \cite{sin_csa_2021}, in the regime of low electrostatic potentials, the nonlinear and linear Poisson-Boltzmann equations predict similar results, whereas in the regime of high electrostatic potentials, the difference in electrostatic potential is very high.
\section{Results and Discussion}
\subsection{Nanochannel grafted with polyzwitterionic brushes}

\begin{figure}
\includegraphics[width=0.7\textwidth]{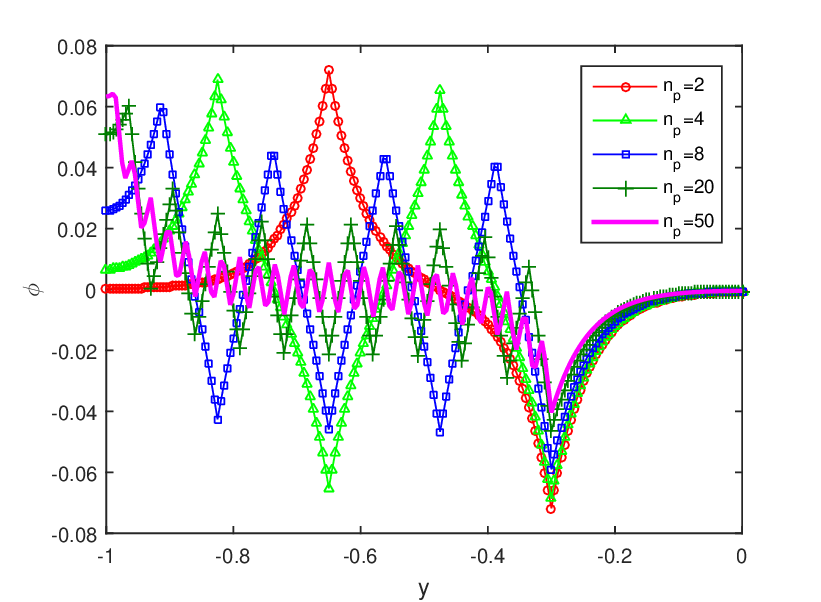}
\caption{ (Color online) Dimensionless electric potential profiles for different charge distributions corresponding to the cases of $n_p=2,4,8,20,50$. Here  each layered charge satisfies  the  following relation $d_k=\frac{0.7k}{n_p}$, $\frac{{ez\sigma _k \lambda H}}{{k_B T\varepsilon _0 \varepsilon }} = \left( { - 1} \right)^{k + 1} ,\left( {k = 1... n_p } \right)$.}
\label{fig:2}
\end{figure}
 
Fig. \ref{fig:2} shows dimensionless electric potential as a function of dimensionless position for a nanochannel grafted with poly-zwitterionic brushes for different counts of layered charges. 

In the condition of $\frac{{ez\sigma _k \lambda H}}{{k_B T\varepsilon _0 \varepsilon }} = \left( { - 1} \right)^{k + 1}$, $\sigma_k$  may be negative or positive, whereas other parameters all are positive. Therefore, the first meaning of this condition is that all the layered charges have an equal magnitude.
The second meaning is that any layered charge with an odd number has positive sign, while any charge with an even number has negative sign. The expression is included in equation of electrostatic potential Eq. (\ref{eq:4} ), for the sake of simplicity, the absolute magnitude is set to 1.
The fluctuation appearing in Fig.\ref{fig:2} ensures discontinuity of electric field strength appearing at locations of layered charges. 
From Fig. \ref{fig:2}, we can confirm that since Debye-H\"uckel equation is a linear differential equation and the principle of superposition is satisfied, the overall self-consistent electric field is the sum of self-consistent electric potentials due to layered charges. Consequently, we insist that when increasing the number of layered charges, variation width of electric potential over the whole space becomes narrower.

\begin{figure}
\includegraphics[width=0.9\textwidth]{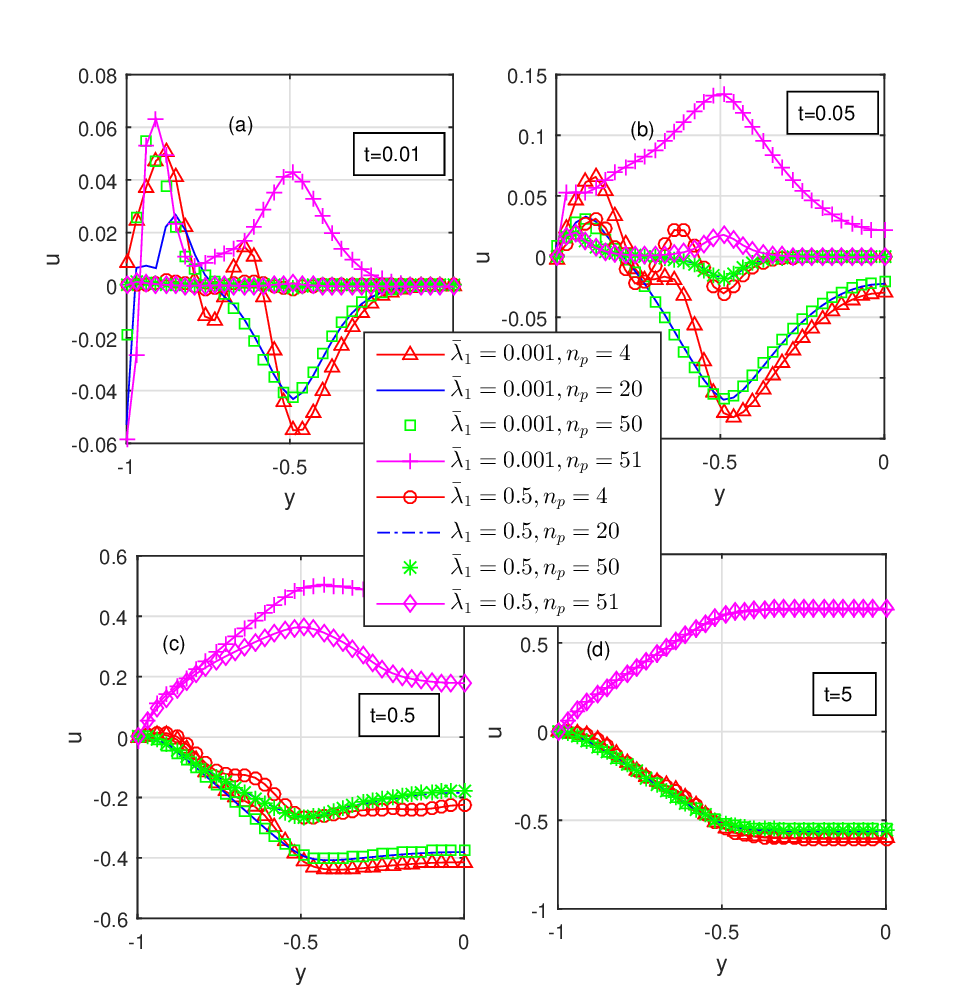}
\caption{ (Color online) Dimensionless electroosmotic velocity for different charge distribution $n_p=4,20,50,51$ and different values of $\lambda_1$ at dimensionless time (a) $t=0.01$, (b) $t=0.05$,(c) $t=0.5$, (d) $t=5$. Here $d=0.5,  \bar \mu _{\rm{c}}=0.2$. Each layered charge satisfies  the  following relation $d_k=\frac{0.5k}{n_p}$, $\frac{{ez\sigma _k \lambda H}}{{k_B T\varepsilon _0 \varepsilon }} = \left( { - 1} \right)^{k + 1} ,\left( {k = 1... n_p } \right)$}
\label{fig:3}
\end{figure}
 
Fig. \ref{fig:3}(a) and (b) demonstrate that for small values of t, velocity profiles exhibit the behavior like the electric potential. This is a natural consequence from linear relation between electric potential and electroosmotic velocity.
In the case of odd $n$, transient electroosmotic velocity is positive, while  if $n$ is even, the velocity has negative values. It can be explained by applying the fact that the charges at distant positions from the surface plays an important role on determination of the direction of the total electroosmotic flow.
Fig. \ref{fig:3}(a)-(d) display that a large $\lambda_1$ requires a more time to reach at steady state as compared to the case of lower $\lambda_1$. This is a consequence from the definition of transient response time.
 
\begin{figure}
\includegraphics[width=0.9\textwidth]{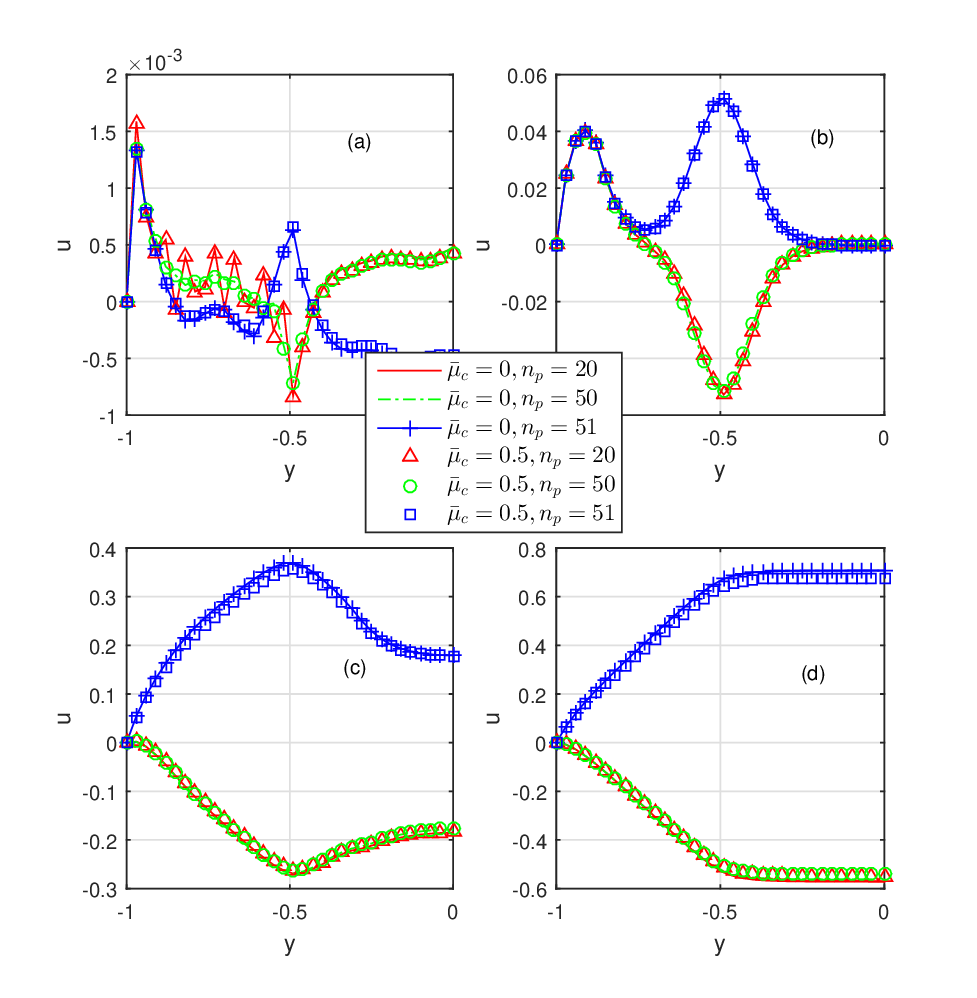}
\caption{ (Color online) Dimensionless electroosmotic velocity as a function of dimensionless location at different values of dimensionless drag coefficient  within polyelectrolyte layer ($\mu_c$) and different charge distribution  $n_p=20,50,51$ for dimensionless time (a) $t=0.01$,(b)  $t=0.05$, (c) $t=0.5$, (d) $t=5$.
 Here $\bar \lambda _1  = 0.5$. Other parameters are the same as in Fig. \ref{fig:3}}
\label{fig:4}
\end{figure}
Fig. \ref{fig:4} depicts that at early time, velocity profile looks like behavior of electric potential distribution. When time flows, depending on the sign of the end-layered charge, the direction of velocity is determined. For cases of even $n_p$, the end layered charge has a negative value, leading to a negative values of electroosmotic velocity. On the other hand, at odd $n_p$, the end layered charge has a positive value and consequently electroosmotic velocity has positive values. Fig. \ref{fig:4}(a)-(d) show that as $\mu_c$ increases, transient and steady velocity are low.  This is a consequence from the fact that a large $\mu_c$ makes a large friction between polyelectrolytes and fluids.
\begin{figure}
\includegraphics[width=0.6\textwidth]{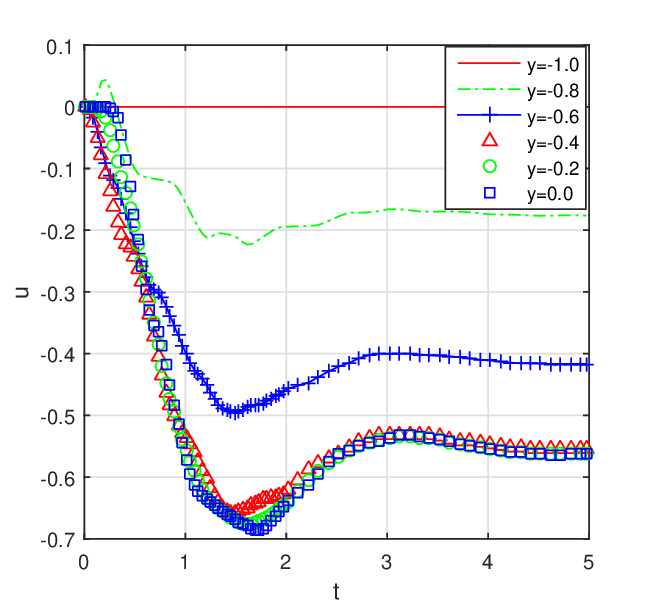}
\caption{ (Color online) Dimensionless electroosmotic flow as a function of dimensionless time at different locations $y=-1, -0.8, -0.6, -0.4, -0.2, 0.0$. $n_p=20$.  Other parameters are the same as in Fig. \ref{fig:4}}
\label{fig:5}
\end{figure} 

Fig. \ref{fig:5} depicts dimensionless electroosmotic flow as a function of dimensionless time at different locations.
Due to viscoelasticity of fluids, through fluctuation processes, velocity reaches at the equilibrium state.

\subsection{Nanochannel grafted with two polyelectrolyte layers of opposite signs}
In this section, we investigate the electroosmotic transport in nanochannel grafted by two polyelectrolyte layers of opposite signs, as shown in Fig. \ref{fig:1}. 
\begin{figure}
\includegraphics[width=0.6\textwidth]{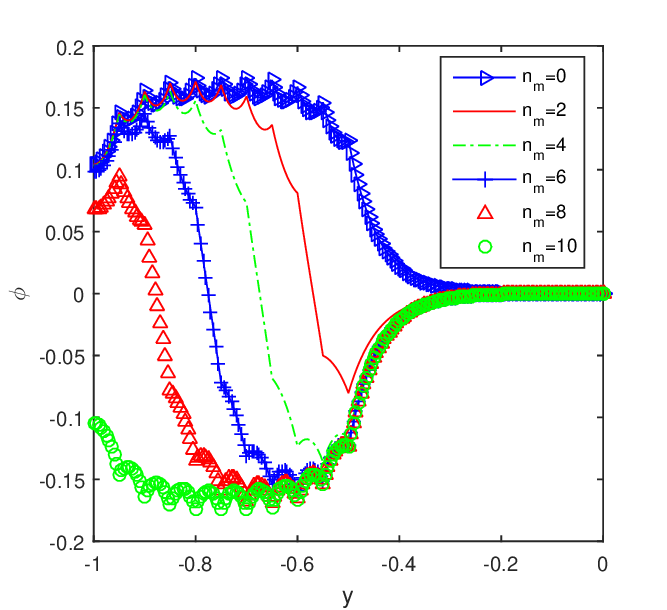}
\caption{ (Color online) Dimensionless electric potential profiles for the cases of $d_k  = \frac{{0.5k}}{{n_p }},k = 1 \sim n_p$
$\sigma _k  = 0.0008c/m^2 (k = 1,2, \cdots ,n_p  - n_m ),\sigma _k  =  - 0.0008c/m^2 (k = n_p  - n_m  + 1, \cdots n_p )$.  
Here $\lambda_1=0.5, n_p=10, d=0.5, H=50nm$.}
\label{fig:6}
\end{figure} 
To relate our study with real-world nanofluidic systems, we have used $\sigma_k= 0.0008C/m^2$ as in \cite{das_2017} and bulk concentration of salt ranging from $0.0001mol/L$ to $0.01mol/L$, as in \cite{khatibi_pf_2022, khatibi_pf_2023, ashrafi_ces_2022}.
 
Fig. \ref{fig:6} shows electric potential profiles for different cases of two polyelectrolyte layers with opposite signs.  The figure displays that since the surface charge densities in one polyelectrolyte layer are equal to each other, the electric potential in the region remains nearly constant and fluctuates slightly.
In the boundary region between two polyelectrolyte layers, electric potential rapidly changes attributed to the fact that charges of opposite sign create electric potential of the opposite sign. 

\begin{figure}
\includegraphics[width=0.5\textwidth]{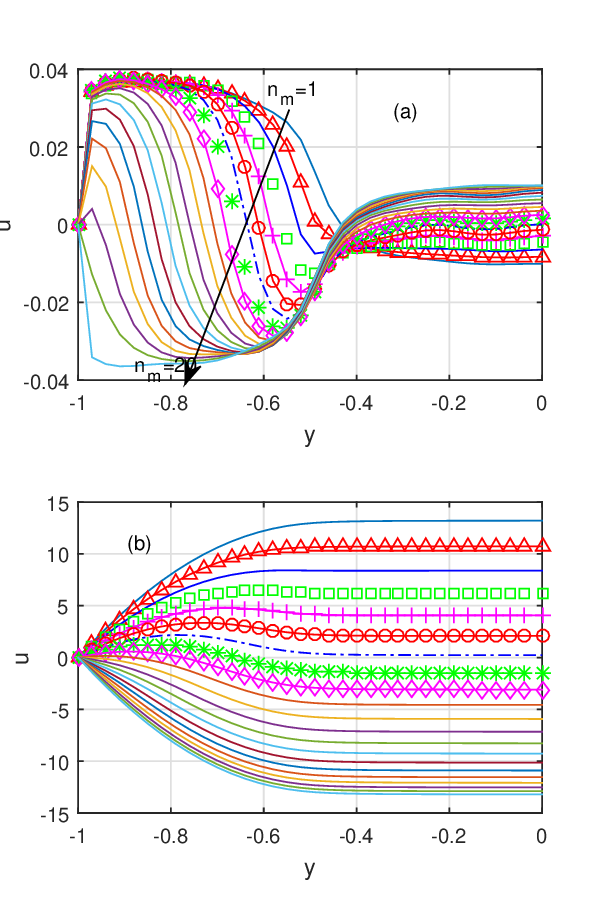}
\caption{ (Color online) Dimensionless electroosmotic velocity profiles for two polyelectrolyte layers grafted nanochannel with different values  of dimensionless time (a) $t=0.05$, (b) $t=5$. $n_p=20$. Other parameters are the same as in Fig. 6.}
\label{fig:7}
\end{figure} 
\begin{figure}
\includegraphics[width=1\textwidth]{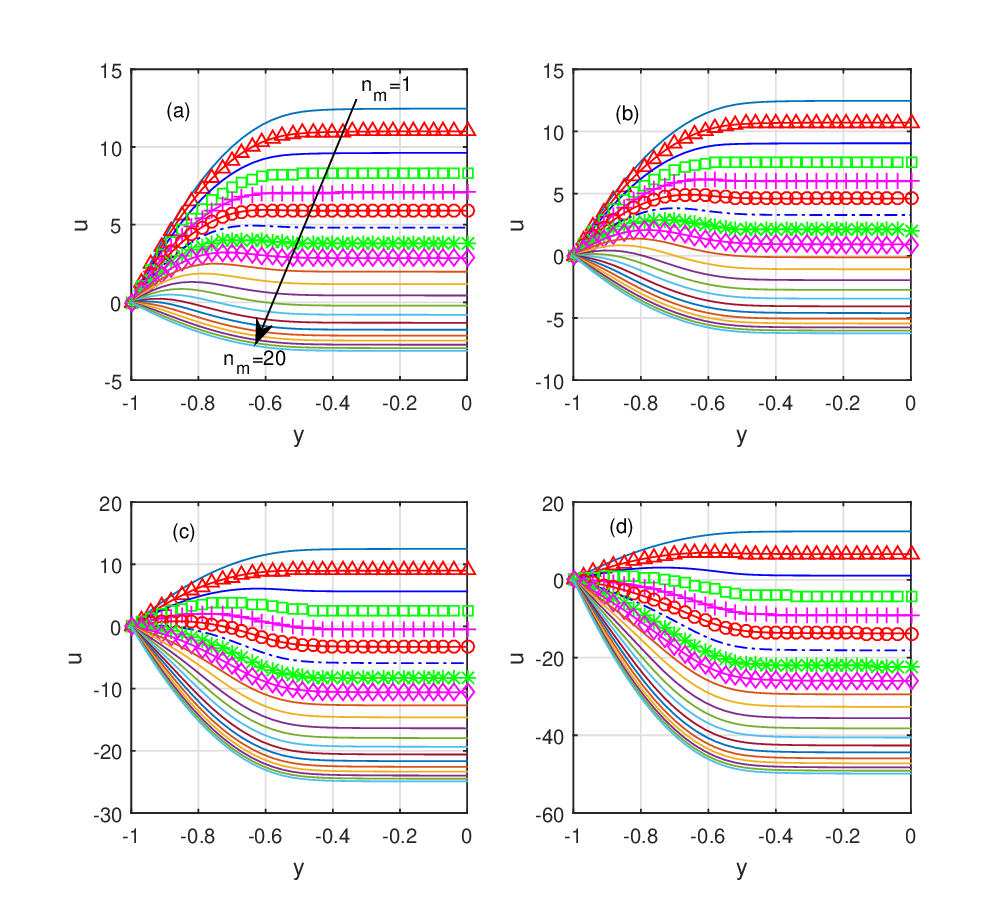}
\caption{ (Color online) Dimensionless electroosmotic velocity profiles for two polyelectrolyte layers grafted nanochannel with different values (a) $k_s=1/4$, (b)$k_s=1/2$, (c)$k_s=2$, (d)$k_s=4$. Here $\bar{\mu}=0.2,\lambda_1=0.5, \sigma_c=0.0008C/m^2, d=0.5, H=50nm, n_p  = 20,d_k  = k/n_p,  n_0=0.01mol/L$. The symbols have the same meanings as in Fig. \ref{fig:7}, respectively}
\label{fig:8}
\end{figure} 

\begin{figure}
\includegraphics[width=0.7\textwidth]{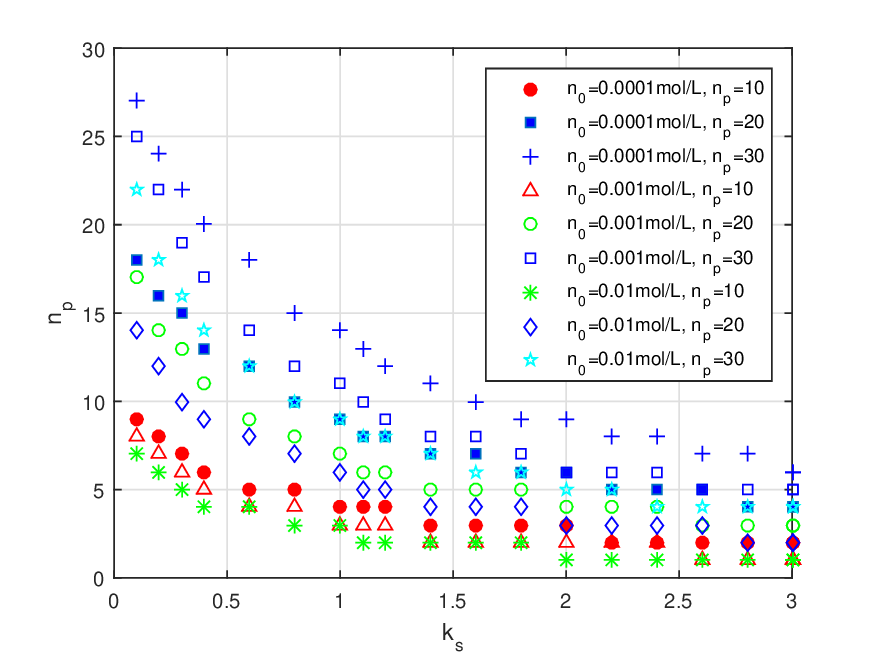}
\caption{ (Color online) Critical number of layered-like charges in polyelectrolyte as a function of $k_s$ for two polyelectrolyte layers grafted nanochannel with different values $n_p=10, 20, 30$. Other parameters are the same as in Fig.6.}
\label{fig:9}
\end{figure}  
 
\begin{figure}[!t]
\begin{tabular}{l}
\includegraphics[clip=true,scale=0.7]{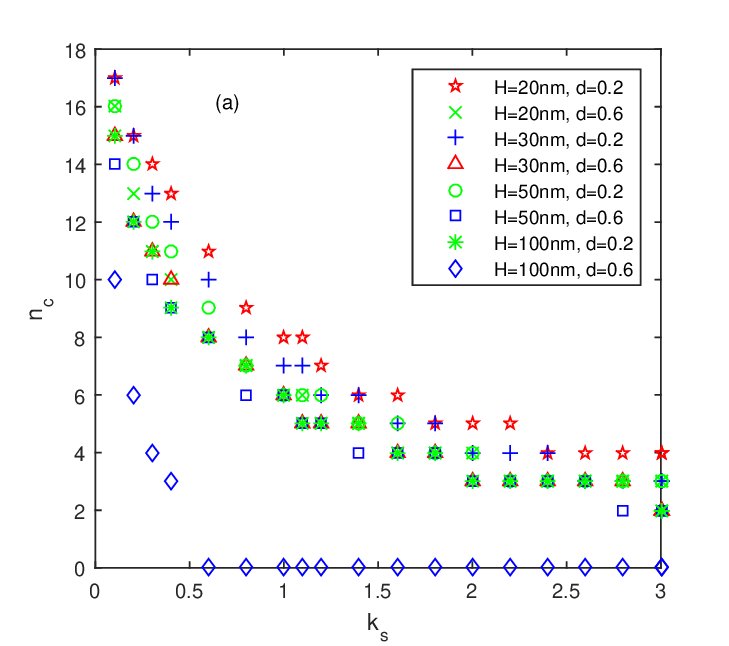}
\includegraphics[clip=true,scale=0.7]{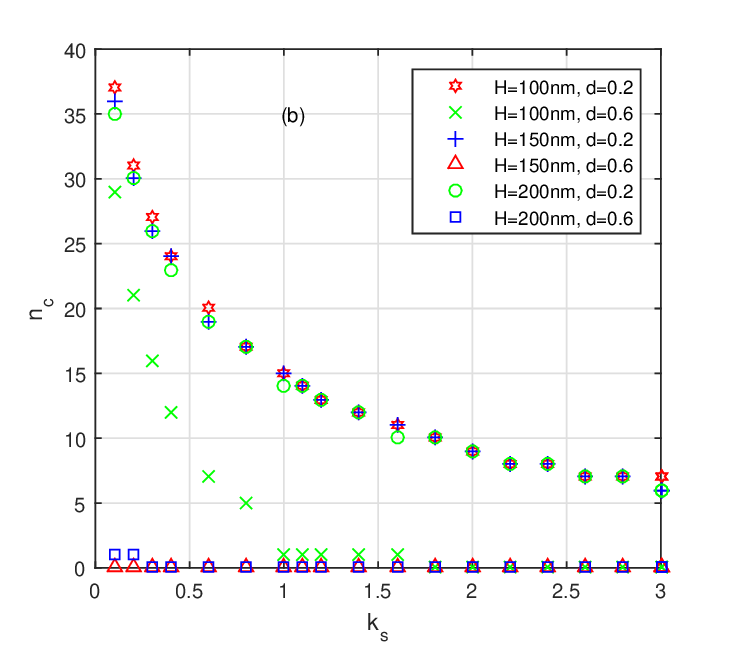}
\end{tabular}
\caption{\label{fig:10} Critical number of layered charges in polyelectrolyte as a function of $k_s$ for two polyelectrolyte layers grafted nanochannel with different values (a) $H=20, 30, 50, 100nm$ and $d=0.2, 0.6, n_p=20$ (b) $H=100, 150, 200nm$ and $d=0.2,0.6.n_p=50$. Other parameters are the same as in Fig.6.}
\end{figure}

Fig. \ref{fig:7} shows dimensionless electroosmotic velocity profiles for two polyelectrolyte layers grafted nanochannel with different numbers of layered charges with opposite sign i.e., $n_m$.
It demonstrates that when thickness of the polyelectrolyte layer of opposite sign becomes large, electroosmotic velocity firstly gets small and then becomes bigger with opposite flow direction.  
Here, we introduce a key factor for quantifying the reversal of electroosmotic flow with respect to a number of layered charges with opposite sign.
The parameter which is called critical number, is defined as: First, we consider a polyelectrolyte layer
with equidistantly distributed, layered charges having an identical charge sign. From the free end of the polyelectrolyte layer toward the nanochannel wall, we one-by-one replace individual layered charge as the opposite signed layered charge. We find that above the number of opposite signed layered charges, the electroosmotic flow at steady state is reversed. We call the number of opposite signed layered charges critical number.

Fig. \ref{fig:8} shows dimensionless electroosmotic velocity for two polyelectrolyte layers grafted nanochannel for different values of the ratio ($k_s$) between layered charges densities in the first and second polyelectrolyte layer.  Assuming that the charge density of each structural layered charges in the first polyelectrolyte layer is  $\sigma_1$, the corresponding one in the second layer with structural charges with opposite sign is $\sigma_2$, $k_s \equiv  -\sigma_2/\sigma_1$.

All of Fig. \ref{fig:8}(a, b, c, d) show that an increase of $k_s$ facilitates the direction of electroosmotic flow to be opposite to original one. 
They also illustrate that a small magnitude of opposite charges requires a longer length of polyelectrolyte layer with charges of opposite sign to reverse flow direction. 
In other words, we can see from Fig. \ref{fig:8}(a)-(d) that the direction of electroosmotic flow is reversed at $n_{c}=12, 8,4,2$, respectively. 

Fig. \ref{fig:9} shows critical number of layered charges as a function of $k_s$ i.e. the ratio ($k_s$) between layered charges densities in the first and second polyelectrolyte layer.
Fig. \ref{fig:9} shows that increasing the layered charge density of the second polyelectrolyte layer yields a decrease of critical numbers of layered charges.  It should be emphasized that an increase of electrolyte concentration yields a small critical number of layered charges. This is attributed to the fact that at a large salt concentration debye screening length becomes small and consequently the first charge layer affects the electroosmotic flow in the central region of a nanochannel weakly. 
  
Fig. \ref{fig:10} shows critical numbers profiles of layered charges in polyelectrolyte as a function of $k_s$ for two polyelectrolyte layers grafted nanochannel. Fig. \ref{fig:10}(a) and (b) shows that for any value of d, a thicker width of nanochannel provides a smaller number of opposite signed layered charges. It is also shown that for any value of H, the critical number for the case of $d=0.6$ is smaller than one for the case of $d=0.2$. In particular, for the case of $d=0.6$ and $H=100nm$, the critical number rapidly gets smaller and even becomes zero. 
All the three phenomena  are attributed to the fact that a layered charge, being at a farther location from nanochannel wall, provides a bigger impact on the electroosmotic flows than others being at closer positions. The theoretical and experimental significance of our approach is that it is applicable to not only uniform charge distribution but also non-uniform charge distribution.

The present study is based on classic Gouy-Chapmann model of electric double layer. In fact, in the regime of high electrostatic potentials, it was already confirmed that the excluded volume of ions and their static polizability has a significant effect on electric double layer. \cite{Kumar_2022, Budkov_2021}  Fortunately, in the regime involved Debye-H\"uckel approximation, such effects are negligible and thus Gouy-Chapmann model remains valid. 

Recently, mean-field theories \cite{Blossey_2017, Budkov_2023} of electrolytes in nanoconfined structures were developed by taking into account structural interactions which not only represent non-spherical shapes of ions, their conformational lability, and solvent effects but also provide significant impact on electosmotic flow.

Our future research will focus on development of a more advanced approach taking into account significant effects, contributing to design of microchip with nanochannels for rectifying electrolyte ions or generating electric power.

\section{Conclusion}

We semi-analytically studied transient electroosmotic flow of general Maxwell fluids through microchannels grafted with multilayers of strong polyelectrolyte brushes with the help of the method of Laplace transform. 

For start-up flow, at initial time, electroosmotic velocity has similar behavior as in electric potential, as time advances, superposition of velocity fields yields small velocity values close to the wall, a larger velocity at a farther distance from the wall, after all, around center position of the nanochannel the velocity is saturated. 

The key result of the paper is controllability of the direction of electroosmotic flow depending on conformation and charge distribution of the layer of polyelectrolyte brushes.
We have elucidated that although mean charge density of polyelectrolyte chains is zero, i.e., the sum of positive and negative structural charges is zero, total electroosmotic flow is finite. Moreover, in such cases, depending on charge distribution at end of polyelectrolyte layers, the direction of electroosmotic flow can be reversed sensitively.

Also, when transient response time becomes larger, the time for reaching at equilibrium state gets longer. Moreover, we ensure that the higher $\mu_c$, the smaller the velocity of electroosmotic flows. 

 We expect that the present methodology will be a powerful tool for studying transient electroosmotic flows in a nanochannel grafted with polyelectrolyte chains having non-uniform charge distribution.

\section{Conflict of interest}
There are no conflicts to declare.

\section{\bf Data Availavility}
The data that support the findings of this study are available from the corresponding author upon reasonable request.

\end{document}